# Direct measurements of DOCO isomers in the kinetics of OD+CO

**Authors:** T.Q. Bui,[1]* B.J. Bjork,[1†] P.B. Changala,[1] T.L. Nguyen,[2] J.F. Stanton,[2] M. Okumura,[3] J. Ye[1]*

**Affiliations:**

[1]JILA, National Institute of Standards and Technology and University of Colorado, Department of Physics, University of Colorado, Boulder, CO 80309, USA

[2]Department of Chemistry, University of Florida, Gainesville, Florida 32611, USA

[3]Arthur Amos Noyes Laboratory of Chemical Physics, California Institute of Technology, Pasadena, California, 91125, USA

[†]Present address: Honeywell International, 303 Technology Ct., Broomfield, Colorado, 80021, USA

*Correspondence to:  thbu8553@jila.colorado.edu, ye@jila.colorado.edu

**Abstract:**

Quantitative and mechanistically-detailed kinetics of the reaction of hydroxyl radical (OH) with carbon monoxide (CO) have been a longstanding goal of contemporary chemical kinetics. This fundamental prototype reaction plays an important role in atmospheric and combustion chemistry, motivating studies for accurate determination of the reaction rate coefficient and its pressure and temperature dependence at thermal reaction conditions. This intricate dependence can be traced directly to details of the underlying dynamics (formation, isomerization, and dissociation) involving the reactive intermediates *cis*- and *trans*-HOCO, which can only be observed transiently. Using time-resolved frequency comb spectroscopy, comprehensive mechanistic elucidation of the kinetics of the isotopic analogue deuteroxyl radical (OD) with CO has been realized. By monitoring the concentrations of reactants, intermediates, and products in real-time, the branching and isomerization kinetics, and absolute yields of all species in the OD+CO reaction are quantified as a function of pressure and collision partner.

**One Sentence Summary:** Frequency comb spectroscopy captures real-time dynamics of *cis*- and *trans*-DOCO isomers produced from the OD+CO reaction.

# MAIN TEXT

# Introduction

Fundamental understanding and control of bimolecular reactions have long been the primary motivations for the fields of chemical kinetics and dynamics. One approach has been to construct a detailed dynamical understanding from the bottom up, exemplified by the field of crossed molecular beams experiments where reaction dynamics are dissected from single collision processes (*1*). For applications in combustion, atmospheric, and fundamental chemistry, a broader goal from these studies is the prediction of the rates and outcomes of reactions. Such knowledge requires direct connection of reaction dynamics to molecular structure and detailed understanding on how the dynamics change during the chemical transformation from reactants to products. However, transitions from reactants to products in bimolecular reactions are often multistep because of the involvement of multiple reactive intermediates, transition states, and final product channels on a multidimensional potential energy surface. These factors have limited experimental characterization of the underlying dynamics, and theoretical models are often required to assist with detailed mechanistic interpretation (*2-4*). It is thus of particular interest to develop new experimental approaches for the study of reaction kinetics and dynamics from real-time observation of individual elementary chemical reactions.

The reaction of hydroxyl (OH) with carbon monoxide (CO), Eq. 1, exemplifies such a system that possesses an intricate reaction mechanism governed by multiple bound states, both reactive and non-reactive:

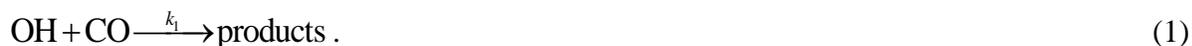

$$OH + CO \xrightarrow{\ k_1\ } products \ . \tag{1}$$

The gas phase kinetics of OH+CO has received considerable attention from both experimental and theoretical efforts due to its fundamental and practical importance in atmospheric and combustion chemistry (*5*). The observed rate coefficient, $k_1$, displays strong pressure and non-Arrhenius temperature dependence, which is now understood to arise from the formation of the HOCO complex, with subsequent tunneling to the products H+CO$_2$. Figure 1 shows the potential energy surface (PES) for the OH+CO reaction.

Over the past four decades, the HOCO formation mechanism had been derived from fits of the observed $k_1(P,T)$ to theoretical models (*6-17*). These studies led to the following proposed reaction mechanism for OH+CO, starting from the calculated asymptotic (stationary) states in Fig. 1:

$$OH + CO \rightleftarrows HOCO^* \rightarrow H + CO_2 . \qquad (2)$$

$$HOCO^* \underset{\longleftarrow}{\overset{[M]}{\longrightarrow}} HOCO . \qquad (3)$$

$$trans\text{-}HOCO \underset{\longleftarrow}{\overset{[M]}{\longrightarrow}} cis\text{-}HOCO . \qquad (4)$$

$$cis\text{-}HOCO \rightarrow H + CO_2 \qquad (5)$$

Reaction 1 initially produces energized HOCO*, which either back-reacts to OH+CO or proceeds to form H+CO$_2$ (Eq. 2). In the presence of a buffer gas, M, HOCO* may be collisionally stabilized (Eq. 3). At low pressures, the overall reaction rate can be represented by an effective bimolecular rate coefficient $k_1([M],T) = k_{1a}([M],T) + k_{1b}([M],T)$, where $k_{1a}$ and $k_{1b}$ describe the formation of stabilized HOCO and H+CO$_2$, respectively.

HOCO exists as two geometrical isomers, *trans*-HOCO and *cis*-HOCO, with the latter being slightly less stable by 5 to 7 kJ mol$^{-1}$ (*18, 19*). Current theory predicts that both isomers will be formed by stabilization upon associative recombination of OH+CO, with the *trans*-HOCO isomer favored. Calculations indicate that the barrier for isomerization (TS4, Fig. 1) is low, leading to rapid isomerization between the two isomers (Eq. 4). However, further reaction to products must proceed through the *cis*-HOCO isomer (Eq. 5), because the transition state to form H+CO$_2$ products from *trans*-HOCO (TS3) is predicted to be much higher than that of *cis*-HOCO (TS2). The *cis*-isomer thus plays a critical role in the OH+CO reaction, and the isomerization dynamics determines product formation. Direct and simultaneous detection of reactant OH, intermediates *cis*-HOCO and *trans*-HOCO, and product H+CO$_2$ and measurement of their time dependence in the same reaction environment would reveal the coupled dynamics of all participating species from beginning to end.

Recently, Bjork *et al.* (*20*) observed the deuterated analogue *trans*-DOCO formed from OD+CO reaction under thermal conditions, providing the first direct experimental confirmation of the HOCO mechanism; however, a full understanding of the reaction mechanism requires observing the *cis*-HOCO radical as well as the final products. While the *cis*-HOCO radical has been detected using molecular or ion beams (*19, 21, 22*) and matrix (*23, 24*) techniques that can trap metastable species, it has not yet been observed in a thermal reaction environment as the product of the OH+CO, leaving a substantial gap in our understanding of this reaction. Because experimental characterization of the OH+CO reaction kinetics is currently fragmented, the state of knowledge for OH+CO mechanism continues to rely on theory (*3, 6, 8, 18, 25*).

We now report a direct observation and measurement of the deuterated analogue *cis*-DOCO intermediate from the OD+CO reaction via time-resolved frequency comb spectroscopy, a technique for obtaining high-resolution, transient absorption spectra over a wide spectral range in the mid-infrared ($\lambda$ = 3 to 5 $\mu$m) with microsecond time resolution (*20, 26-28*). Using a room-temperature flow cell, the OD+CO reaction is initiated at $t = 0$ by photo-dissociation of $O_3$ in the presence of $D_2$, CO, and $N_2$ (total pressure range of 20 to 100 torr). From real-time measurements of the time-dependent populations of reactant OD, intermediates *trans*-DOCO and *cis*-DOCO, and product $CO_2$ via their infrared absorption spectral windows shown in Fig. 2, we disentangle the entire kinetics process by following this reaction step-by-step. We now have direct access to the reaction rate coefficients and branching yields of all chemically accessible product channels; this provides the most comprehensive experimental observations to date of this multistep reaction.

**Results and Discussion**

Figure 2A shows the gas phase spectrum of *cis*-DOCO in the OD stretch fundamental band, which has not been observed prior to this work. To confirm the identification of *cis*-DOCO, we perform additional experiments to obtain an isotopic substitution spectrum. The predominately *b*-type nature of the *cis*-DOCO OD stretch band gives rise to an observed progression of Q-branches, each consisting of many unresolved individual rovibrational transitions, spaced approximately 7 cm⁻¹ apart (Table S1). By fitting the frequencies of these spectral features, the $v$=1 band origin and $A$ rotational constant for both ¹²C and ¹³C isotopologues of *cis*-DOCO are obtained. The predicted and measured spectroscopic constants for both ¹²C and ¹³C isotopologues of *cis*-DOCO are compiled in Table S2. The measured isotopic shifts to these molecular constants are then compared to their respective *ab initio* values predicted using second-order vibrational perturbation theory (VPT2) at the CCSD(T)/ANO1 level (Table S3). The agreement between the experimental and theoretical values provides conclusive evidence for the identification of *cis*-DOCO.

The time-dependent concentrations of OD, *cis*-DOCO, *trans*-DOCO (Fig. 2B), and $CO_2$ are determined from the infrared absorption signals; time traces at a single reaction condition are plotted in Fig. 3A. While the reaction system is characterized by a complex set of primary and secondary reactions, as described in our earlier work (*20, 28*), we find that at early times ($t < 200$ μs) we can effectively model the time dependence including only formation, isomerization, and loss processes of DOCO isomers, represented by rate coefficients $k_{1a,i}([M],T)$, $k_{iso,ij}([M],T)$, and $k_{loss,i}$ ($i,j = cis$ or $trans$), respectively, with the following rate equations (Eq. 6):

$$\frac{\mathrm{d}[cis]}{dt} = k_{1a,cis}[CO][OD]_t - k_{loss,cis}[X][cis]_t - k_{iso,ct}[cis]_t + k_{iso,tc}[trans]_t$$
$$\frac{\mathrm{d}[trans]}{dt} = k_{1a,trans}[CO][OD]_t - k_{loss,trans}[X][trans]_t + k_{iso,ct}[cis]_t - k_{iso,tc}[trans]_t \qquad (6)$$

The subscript $t$ denotes time. The time-dependent [$cis$-DOCO]$_t$ and [$trans$-DOCO]$_t$ are denoted by [$cis$]$_t$ and [$trans$]$_t$, respectively. $k_{iso,ct}([M],T)$ and $k_{iso,tc}([M],T)$ are the pressure-dependent effective isomerization rate coefficients for $cis{\rightarrow}trans$ and $trans{\rightarrow}cis$, respectively. $k_{loss}$ describes the decay of DOCO from a reaction with one or more species X, and [OD]$_t$ is the time-dependent concentration of OD in the ground vibrational state. We have previous characterized vibrationally excited OD populations and lifetimes at our experimental conditions; nonequilibrium effects introduce no more than a 10% error in our current measurements (*20*). We assume that the dissociation/tunneling of thermalized $cis$-DOCO to D+CO$_2$, which proceeds over a high barrier (TS2), cannot compete with the other channels at these timescales.

Considering the possibility of substantial isomerization rates that may compete with the formation rate, we set out to perform experiments to disentangle their relative effects by varying the buffer gas species and its pressure. We sought to distinguish formation from isomerization by first determining the effective bimolecular formation and loss rate coefficients for the sum of DOCO isomers, $k_{1a,sum}([M],T) \equiv k_{1a,cis}([M],T)+k_{1a,trans}([M],T)$ and a similar expression for $k_{loss,sum}$ (Eq. 7).

$$\frac{\mathrm{d}[DOCO_{sum}]}{dt} = k_{1a,sum}[CO][OD]_t - k_{loss,sum}[X][DOCO_{sum}]_t . \qquad (7)$$

Here, [DOCO$_{sum}$] $\equiv$ [$trans$]+[$cis$]. Using the total DOCO concentration eliminates the need to model isomerization in Eq. 6 (derived in Section 2 of the Supplementary Materials (SM)). In the low-pressure limit studied here, the DOCO formation rate obeys a termolecular rate law in which M represents the buffer gas species (a third-body collider). In this reaction, M comprises all the high density species present: $N_2$, CO, $D_2$, and $O_3$.

The solution to Eq. 7 is a convolution of the DOCO loss term with $[OD]_t$, given by the integral in Eq. 8 ($u$ is a dummy variable). [CO] is in large excess and remains effectively constant throughout the reaction.

$$[DOCO_{sum}]_t = k_{1a,sum}[CO]\int_0^t e^{-(k_{loss,sum}[X])(t-u)}[OD](u)du .$$  (8)

The effective bimolecular rate coefficient $k_{1a,sum}([M],T)$ can be decomposed into three terms dependent on $N_2$, CO, and $D_2$ concentrations,

$$k_{1a,sum} = k_{1a,sum}^{(CO)}[CO] + k_{1a,sum}^{(N2)}[N_2] + k_{1a,sum}^{(D2)}[D_2] + k_{1a,sum}^{(0)},$$  (9)

where $k_{1a,sum}^{(CO)}$, $k_{1a,sum}^{(N2)}$ and $k_{1a,sum}^{(D2)}$ are the termolecular rate coefficient dependence on M=CO, $N_2$ and $D_2$, respectively. $k_{1a,sum}^{(0)}$ is an offset term that accounts for additional background species. $O_3$ is not included in Eq. 9 because its concentration is smaller by factor greater than $10^2$ compared to the other buffer gas species. Its main contribution is to DOCO loss.

By simultaneously fitting $[DOCO_{sum}]_t$ and $[OD]_t$ from data in Fig. 3A across varying [CO], $[N_2]$, and $[D_2]$, the effective bimolecular rate coefficients for DOCO$_{sum}$ formation, $k_{1a,sum}$, are obtained as follows. $[OD]_t$ are fit using empirically derived analytical functions (Eq. S4). This spline interpolation procedure provides an analytical expression for $[OD]_t$ without making any assumptions about its formation or decay mechanisms. The observed $[DOCO_{sum}]_t$ is then fit to Eq. 8, integrating the fitted $[OD]_t$ function over the time window of -25 (convoluted with the finite response time of the camera) to 200 μs. The fitted parameters in Eq. 8 are $k_{1a,sum}$ and $r_{loss,sum}$; the latter is an effective first-order rate for loss of DOCO$_{sum}$.

DOCO must be lost by reaction with one or more species, which we define as X; then, $r_{loss,sum} = k_{loss,sum}[X]$, where $k_{loss,sum}$ is a bimolecular rate coefficient. We hypothesized that X = $O_3$, since $O_3$ is the highest concentration candidate that is reactive with DOCO. To test this, $r_{loss,sum}$ is determined independently by measuring $[cis]_t$ and $[trans]_t$ while varying the density of $O_3$; the results are shown in Fig. S1. The observed loss rate is proportional to $O_3$ and is invariant to [CO], $[N_2]$, and $[D_2]$, leading to an empirical bimolecular rate coefficient for DOCO loss to $O_3$, $k_{loss,sum}$ = 2.5(6)×$10^{-11}$ cm$^3$ molecules$^{-1}$ s$^{-1}$, from a global fit across all CO, $N_2$, and $D_2$ data sets. We conclude that X is predominantly $O_3$. One possible outcome from the reaction of DOCO with $O_3$ is to recycle OD which, together with the reaction of D+$O_3 \rightarrow$ OD+D, would lead to the quasi-steady-state observed in $[OD]_t$ at long delay times (Fig. 3A). The direct measurement of $[OD]_t$ eliminates the need to explicitly treat the recycling in our analysis. Finally, because there is some $O_2$ present in the $O_3$ flow, some of the loss is due to reaction with $O_2$, which will appear to vary linearly with $O_3$ as well. Any contribution to $r_{loss,sum}$ from reaction with $O_2$ is likely to be minor, as $[O_2] << [O_3]$, and additionally, the rate coefficient for $DOCO+O_2$ is an order of magnitude lower than $k_{loss,sum}$.

After establishing that $r_{loss,sum}$ is due to reaction with $O_3$, we next determine the effective bimolecular rate coefficients, $k_{1a,sum}$, to form DOCO from Eq. 8 for each buffer gas ($[CO]$, $[N_2]$ and $[D_2]$). The results from experiments varying each gas concentration independently while fixing $[O_3] = 2\times10^{15}$ molecules cm$^{-3}$ ($r_{loss,sum} = 5\times10^4$ s$^{-1}$), are displayed as data points in Figs. 3B, 3C, and S2, respectively.

Fits of $k_{1a,sum}$ reveal that under our conditions, CO and $N_2$ contribute to DOCO stabilization; there is no observable dependence on $D_2$ over the limited pressure range studied. To account for all three species present in each experiment, we perform a multidimensional linear regression of the data in Figs. 3B, 3C, and S2 (fitted bimolecular rate coefficients) to Eq. 9 in order to determine the termolecular rate coefficients $k_{1a,sum}^{(CO)}$, $k_{1a,sum}^{(N2)}$, and $k_{1a,sum}^{(D2)}$, respectively. The linear fits are given as red dashed lines. In the varying CO experiment (Fig. 3B), a clear linear dependence is observed, indicating a termolecular dependence of $k_{1a,sum}$ on CO, or $k_{1a,sum}^{(CO)}$. The offset in the linear fit comes from the $N_2$ termolecular dependence of $k_{1a,sum}$, or $k_{1a,sum}^{(N2)}$. A similar linear dependence is seen for varying $N_2$ (Fig. 3C), with the offset being the $k_{1a}^{(CO)}$ component. The results obtained from the multidimensional linear regression are $k_{1a,sum}^{(CO)} = 8.5_{-5.4}^{4.5} \times 10^{-33}$ cm$^6$ molecules$^{-2}$ s$^{-1}$, $k_{1a,sum}^{(N2)} = 1.3_{-0.5}^{0.4} \times 10^{-32}$ cm$^6$ molecules$^{-2}$ s$^{-1}$ and $k_{1a,sum}^{(D2)} = 9.6_{-24}^{23} \times 10^{-33}$ cm$^6$ molecules$^{-2}$ s$^{-1}$. We observe an additional offset in the effective bimolecular rate constant, $k_{1a,sum}^{(0)}$ $= 1.5_{-0.8}^{0.7} \times 10^{-14}$ cm$^3$ molecules$^{-1}$ s$^{-1}$. More than 80% of the offset term, $k_{1a,sum}^{(0)}$, can be accounted for by a combination of the upper bound of $k_{1a,sum}^{(D2)}$ and competition from DOCO loss processes ($r_{loss,sum}$). The $D_2$ concentrations are too low (<10% of total gas) to determine $k_{1a,sum}^{(D2)}$ accurately. Discussions of all statistical and systematic error sources for $k_{1a,sum}$ are found in Table S4 of the SM.

Up to this point, we have considered only the formation kinetics of the two DOCO isomers together and thus could neglect *cis/trans* isomerization. To resolve the dynamical coupling of the isomer-specific formation and isomerization kinetics, we measure the time dependence of the [*trans*]/[*cis*] ratio at nine different partial pressures of CO. Fig. 4A shows five representative experimental time traces of this ratio, which are offset vertically for reasons of clarity. The observed time evolution of the [*trans*]/[*cis*] ratio to a steady-state value for each trace indicates a transition between two distinct time regimes: the early-time DOCO formation ($k_{1a}$) and subsequent relaxation by *trans/cis* isomerization ($k_{iso}$).

We hypothesize that the observed time dependence of the [*trans*]/[*cis*] ratios can be described by a simple kinetics model given by Eqs. S9 to S17 and qualitatively understood as follows. The nascent [*trans*]/[*cis*] ratio extrapolated to $t \approx 0$ μs (prior to any thermal, canonical isomerization) is determined only by the ratio of the formation rate coefficients for the two isomers ($k_{1a,trans} / k_{1a,cis}$). This initial ratio then rises/decays to a steady-state value. The rate of the exponential rise/decay increases with sum of $k_{iso,tc}$ and $k_{iso,ct}$ and is an indication of how fast the *trans* to *cis* populations approach this steady-state. During this time ($t < 50$ μs), formation and isomerization compete. At $t > 50$ μs, the isomerization reaction dominates and establishes steady-state populations of *trans* and *cis* isomers that persist at long times. If the formation and loss processes for DOCO are slow relative to isomerization, the *trans* and *cis* isomers are in equilibrium. The [*trans*]/[*cis*] ratio then becomes independent of [CO] and [*trans*]/[*cis*] = $k_{iso,ct} / k_{iso,tc} = K_{iso}$, the equilibrium constant ($K_{iso}$) for DOCO isomerization.

To quantify the isomerization and isomer-specific-formation rate coefficients, we fit time-dependent [*trans*]/[*cis*] ratios by extending Eq. 6 to include the rate equation for OD formation and decay (Eq. S7). The dashed lines ($\chi^2_{red} \approx 0.9$ to $1.7$) in Fig. 4A are results obtained from fitting the formation rate coefficients, $k_{1a,cis}$ and $k_{1a,trans}$, and isomerization coefficients, $k_{iso,tc}$, and $k_{iso,ct}$, to all nine experimental data sets (differing in CO pressures) simultaneously. Here, $k_{1a,sum}=k_{1a,cis}+k_{1a,trans}$ is fixed to the value measured previously for the sum of isomers (in Fig. 3B). We also assume that the loss rate is isomer-independent and fix $k_{loss}$ to the fitted value in Eq. 8.

The CO dependence of the fitted isomer specific $k_{iso}^{fit}$ and $k_{1a}$ are given in Figs. 4B and 4C, respectively, and compiled in Table S5. Both the formation and thermalized isomerization rate constants are observed to be linear with pressure and hence in the low-pressure limit. For the termolecular formation rate coefficients, we find $k_{1a,trans}^{(CO)} = 1.4(4)\times10^{-32}$ cm$^6$ molecules$^{-2}$ s$^{-1}$ and $k_{1a,cis}^{(CO)} = 6(2)\times10^{-33}$ cm$^6$ molecules$^{-2}$ s$^{-1}$ from fitting the data in Fig. 4C. The ratio of the fitted isomerization rate coefficients (inset, Figure 4B), however, varies with CO concentration, and only becomes constant at higher [CO]. This ratio should be a constant and independent of [CO]; this variation, which reflects different values of the steady-state [*trans*]/[*cis*] ratio, indicates the existence of competition between DOCO isomerization, bimolecular formation, and reactive loss. The fitted isomerization rates are thus empirical, and we designate them as apparent isomerization rate coefficients $k_{iso,tc}^{fit}$ and $k_{iso,ct}^{fit}$.

Figure 4A show that with increasing [CO], the transition from decay to rise behavior in the [*trans*]/[*cis*] ratio occurs at approximately [CO] = $6.5\times10^{17}$ molecules cm$^{-3}$. This trend can now be explained by comparing the $k_{iso,ct}^{fit} / k_{iso,tc}^{fit}$ and $k_{1a,trans} / k_{1a,cis}$ ratios given in the insets of Figs. 4B and 4C, respectively. This transition is also the intersection point of the decreasing $k_{iso,ct}^{fit} / k_{iso,tc}^{fit}$ and increasing $k_{1a,trans} / k_{1a,cis}$. At the lowest CO pressures, the formation, isomerization and loss rates all have similar magnitudes ($\approx 10^4$ s$^{-1}$), and therefore these processes compete.

Additional evidence for this interplay can be found in the observed ratio of $k_{iso,ct}^{fit} / k_{iso,tc}^{fit}$ from the inset of Fig. 4B. Under these conditions, the fitted ratio of $k_{iso,ct}^{fit} / k_{iso,tc}^{fit}$ varies apparently with CO pressure, indicating that these empirical parameters in our model include isomer-specific formation and loss processes. However, at higher pressures ([CO] > $6.4\times10^{17}$ molecules cm$^{-3}$), isomerization dominates ($\approx 10^5$ s$^{-1}$) and $k_{iso,ct}^{fit} / k_{iso,tc}^{fit}$ converges to a constant value, which suggests that the high pressure ratio is the true ratio of isomerization rates and hence the pressure-independent equilibrium constant, $K_{iso} \approx 5(2):1$.

The *trans*- and *cis*-DOCO equilibrium constant is a fundamental thermodynamic quantity. Our measured equilibrium constant is consistent with that determined from the experimental well depths of HOCO isomers reported by Continetti and coworkers (*19*). We can also compare our measurement with predictions by state-of-the-art calculations. Using a theoretical method that combines semi-classical transition state theory (*29-31*) with 2-dimensional master equations (*18, 25, 32*), the calculated pressure-dependent isomerization rate coefficients are displayed in Fig. S3.

The theoretical equilibrium constant is $K_{iso,th} \approx 14{:}1$. While the measured value disagrees with theory, a change of $\approx 2$ kJ/mol in the relative energies (enthalpy) of *cis*- and *trans*-DOCO would account for the difference in the measured and predicted equilibrium constants. Since the entropy component of the $K_{eq}$ calculation comes from the use of theoretical ro-vibrational parameters that are in good agreement with the experimentally measured values, it is unlikely that entropy accounts for a majority of the discrepancy. Our observed equilibrium concentrations also rely on *ab initio* calculations of the absorption cross sections of *cis*- and *trans*-DOCO, which are challenging for accurate determination.

New experimental insight into the energy transfer processes is obtained from the formation rate dependence on collision partner and total pressure for both *cis*- and *trans*-DOCO isomers. We demonstrate that DOCO exhibits isomer-specific formation rates. CO is more than twice as effective in collisionally stabilizing *trans*-DOCO* compared to *cis*-DOCO*. The unusually large enhancement by the CO collision partner for the *trans*-DOCO isomer relative to *cis*-DOCO may be a consequence of complex formation between *trans*-DOCO* and CO from a stronger interaction potential, which would facilitate energy transfer. This mechanism is further supported by spectroscopic observation of the *trans*-HOCO(CO) complex in the molecular beam environment by Oyama *et al.* (*33*).

Finally, the identification and quantification of *cis*-DOCO, *trans*-DOCO, and D+$CO_2$ accounts for all possible branching pathways for OD+CO. It is now experimentally possible to deconstruct the effective bimolecular rate coefficient $k_1$ in the low pressure limit as a sum of elementary reaction steps, i.e. $k_1([M],T) = k_{1a,cis}([M],T) + k_{1a,trans}([M],T) + k_{1b}(T)$, where $k_{1b}(T)$ is the bimolecular rate coefficient for the pressure-independent well skipping D+$CO_2$ channel (*6, 7, 28*). This approximation is valid in the low pressure limit, provided $k_1$ is only strongly dependent on $k_{1a}$ while $k_{1b}$ remains constant with total pressure and collision partner. The latter is independently verified from our measurements of the pressure-dependent formation rate of the $CO_2$ (*28*). Previously, $k_1$ was determined only with contributions from *trans*-DOCO and $CO_2$. Mass balance dictates that *cis*-DOCO also contributes to the total $k_1$; the pressure dependence of $k_1$ has been plotted in Fig. S4. The observed agreement in $k_1$ with previous studies by Golden *et al.* (*8*) and Paraskevopoulos and Irwin (*11*) provides additional confirmation that $k_1$ can be treated as a sum of the rate coefficients of independent channels, $k_{1a}$ and $k_{1b}$, in the low-pressure limit. The $k_{1b}$ value measured previously (*28*) reproduces the measured $[CO_2]_t$ in this work (Fig. 3A, dotted line fit to Eq. S8), which provides additional confirmation of the OD+CO → D+$CO_2$ channel. We note that at $t > 300$ μs, [$CO_2$] significantly exceeds [OD] because (1) there are no known reactive channels causing $CO_2$ loss and (2) secondary reactions that produce $CO_2$ (e.g. DOCO+$O_3$ → OD+$CO_2$+$O_2$) result in its accumulated concentration.

**Conclusions**

We have unambiguously identified and quantified the time-dependent concentrations of all the reactant, intermediate, and product channels in the direct kinetics of OD+CO using time-resolved frequency comb spectroscopy. Each step of the reaction was analyzed to obtain detailed information about the elementary chemical reactions and product branching ratios. How each step quantitatively contributes to the apparent reaction rate coefficient $k_1$ – the main observable for previous studies – is now experimentally confirmed. In addition, new insights are obtained into both the reaction mechanism and the energy transfer processes from the observed differences in the formation rate and reactivity of DOCO isomers. For the first time, these experimental observations capture DOCO dynamics where formation of DOCO is directly followed by isomerization reactions between the *cis*- and *trans*-DOCO isomers. Typically, isomers of intermediate species are observed either in equilibrium or isolated (*34, 35*): to see the transition in this intermediate regime is unusual and highlights the richness of the reaction landscape. Finally, we anticipate that the next exciting frontier for the OH+CO reaction is in the low-temperature regime where tunneling is particularly conspicuous. With the rapid emergence and success of cold molecule physics (*36*), the OH+CO reaction provides an ideal system to learn new physics and chemistry ranging from cold to the ultra-cold regime, especially from the perspective of quantum control of chemical reactions.

**Materials and Methods**

**Time-resolved frequency comb spectroscopy**

The OD+CO reaction is initiated in a room temperature ($T \approx 295$ K) flow cell enclosed by high reflectivity mirrors (for cavity-enhanced absorption spectroscopy). Absorption spectra are obtained from a mid-IR frequency comb (probe beam) propagating collinearly with the cavity axis. Details for the frequency comb source, enhancement cavity, and dispersive spectrometer can be found in previous reports (*20, 27*). UV light at 266 nm ($\approx$ 30 mJ/pulse, $\approx$ 10 ns) orthogonal to the cavity axis photo-dissociates $O_3$ at $t = 0$, which produces $O(^1D)$ and $O_2$ at near unity yield. Typically, about 15% of the $O_3$ is dissociated per pulse. In the presence of $D_2$ gas, the reaction $O(^1D)+D_2$ produces $OD(v = 0$ to $4)$ and D atoms. $OD(v > 0)$ populations are rapidly quenched to the vibrational ground state at our experimental conditions. $OD(v \geq 0)$ internal state distributions are thermalized at $t > 10$ μs (*20*). Addition of CO gas initiates the OD+CO reaction. The total pressure range for all experiments is 20 to 100 Torr. The reaction kinetics are obtained by setting a variable time delay between the photolysis pulse and the IR probe beam acquisition.

Several key experimental modifications have led to an improvement of the measurement signal-to-noise ratio compared to our previous work (*20*). The absorption sensitivity and spectral coverage are improved by the use of high finesse mirrors with a larger bandwidth ($\approx 3.6$ to $4.2$ μm). This broad spectral span grants access to a larger number of species formed from OD+CO, including $OD(v \geq 0)$, $DO_2$, $CO_2$, *cis*-DOCO, and *trans*-DOCO. To observe each species, we tune the OPO wavelength to its respective spectral windows. The 266 nm photolysis beam profile is shaped to be 130 mm x 6 mm to optimize the production of DOCO and spatial overlap with the mid-IR probe beam. We also utilize a retro-reflector mirror for the photolysis beam to double the photolysis fraction. These combined changes have resulted in a net three-fold increase in the measured signal of *trans*-DOCO relative to the previous apparatus.

**Data extraction and analysis**

*Spectral fitting:* The total intracavity absorbance is fit to a linear sum of absorbances for molecular species A, B, etc.:

$$A(\tilde{v}) = -\log\left(\frac{I_S(\tilde{v})}{I_R(\tilde{v})}\right) = l_{eff}[n_A(t)\sigma_A(\tilde{v}) + n_B(t)\sigma_B(\tilde{v}) + ...], \tag{9}$$

where $A(\tilde{v})$ is the cavity absorbance, $n_p$ ($p = $ A, B, etc.) is the molecular density (molecules cm$^{-3}$), $\sigma_p$ is the molecular absorption cross section (cm$^2$), $l_{eff}$ is the effective absorption path length (cm), and $\tilde{v}$ is wavenumber (cm$^{-1}$). The reference intensity, $I_R(\tilde{v})$, is recorded 4 ms before the photolysis pulse and the signal intensity, $I_S(\tilde{v})$, is recorded after the photolysis pulse at variable delay time $t$.

*Spectral line intensities:* The OD, $D_2O$, and *trans*-DOCO line positions and intensities used in this work have been described previously (*20*). The $CO_2$ line positions and intensities are obtained from HITRAN 2012 (*37*). The *cis*-DOCO spectral parameters compiled in Table S2 are used to simulate the ro-vibrational spectrum in PGopher (*38*). The theoretical *cis*-DOCO $v$=1 band strength of $S_{\text{cis-DOCO}} = 14.4124$ km/mol is calculated using VPT2 at the CCSD(T)/ANO1 level of theory.

*Photolysis Path Length and Finesse:* The effective absorption path length, $l_{\text{eff}}$, is determined from the width of the photolysis beam, $l_{\text{phot}}$, and the cavity finesse, $F$, as

$$l_{eff} = \frac{\beta F l_{phot}}{\pi} \, . \tag{10}$$

Here, $\beta = 1$. The cavity finesse as a function of wavelength is measured using the cavity ringdown technique. From Eq. 10, the effective path length is $331 \pm 26$ m at the peak finesse at 3920 nm. This effective absorption path length as a function of wavelength is used to determine the absolute densities displayed in Fig. 3A. The $k_{1a,\text{sum}}$ measurements, however, are independent of finesse since we calculate the time derivative of [DOCO] normalized to [OD]$_t$, which cancels out the path length contribution.

**Supplementary material for this article is available at**

Fig. S1. Variation of $r_{\text{loss,sum}}$ with $O_3$ concentration.

Fig. S2. Variation of $k_{1a,\text{sum}}$ with $D_2$ concentration.

Fig. S3: Calculated *cis/trans* isomerization rates for DOCO isomers using semi-classical transition state theory (SCTST) and 2D master-equation.

Fig. S4: Measured $k_1$ comparison with previous works.

Table S1: Experimental spectral features for *cis*-DO$^{12}$CO and *cis*-DO$^{13}$CO and their ro-vibrational assignments.

Table S2: Summary of ro-vibrational spectral parameters for *cis*-DO$^{12}$CO and *cis*-DO$^{13}$CO.

Table S3: Comparison of experimental and theoretical isotopic shifts for *cis*-DOCO.

Table S4: Summary of statistical (stat) and systematic (sys) errors for $k_{1a,\text{sum}}$.

Table S5: Summary of fitted rate coefficients.

**Acknowledgements:**

Additional data supporting the conclusions are available in supplementary material. We thank J. M. Bowman and H. Guo for stimulating discussions. We acknowledge financial support from AFOSR, DARPA SCOUT, NIST, and NSF JILA Physics Frontier Center. J.F. Stanton and T. L. Nguyen acknowledge financial support from the U.S. Department of Energy, Office of Basic Energy Sciences for Award DE-FG02-07ER15884. M. Okumura acknowledges support of NSF grant CHE-1413712 and the NASA Upper Atmospheric Research Program UARP. T. Q. Bui is supported by the National Research Council Research Associate Fellowship, P. B. Changala is supported by the NSF GRFP. **Author Contributions:** T.Q.B., B.J.B, M.O., and J.Y. conceived and designed the experiments. T.Q.B., B.J.B., P.B.C., and J.Y. discussed and implemented the experimental technique, and T.Q.B. and B.J.B. analyzed all data. T.L.N and J.F.S. provided supporting theory. All authors contributed to technical discussions and writing of the manuscript. **Competing interests:** All authors declare that they have no competing interests. **Data and materials availability:** Data are archived on a network attached storage drive located at JILA. Data will be made available upon request.

**Figures:**

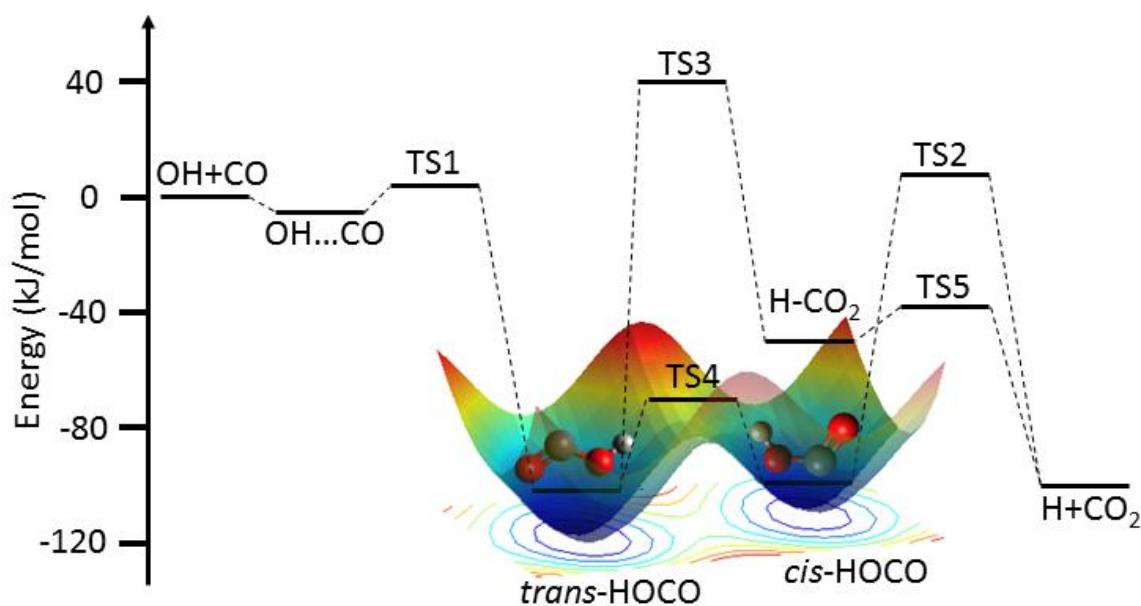

**Fig. 1**: The OH+CO reaction potential energy surface (PES) with energies in units of kJ mol$^{-1}$. The colored PES in the background depicts the HOCO isomer potential energy wells along the H-O-C-O dihedral angle (parallel to page) and O-H bond length (perpendicular to page).

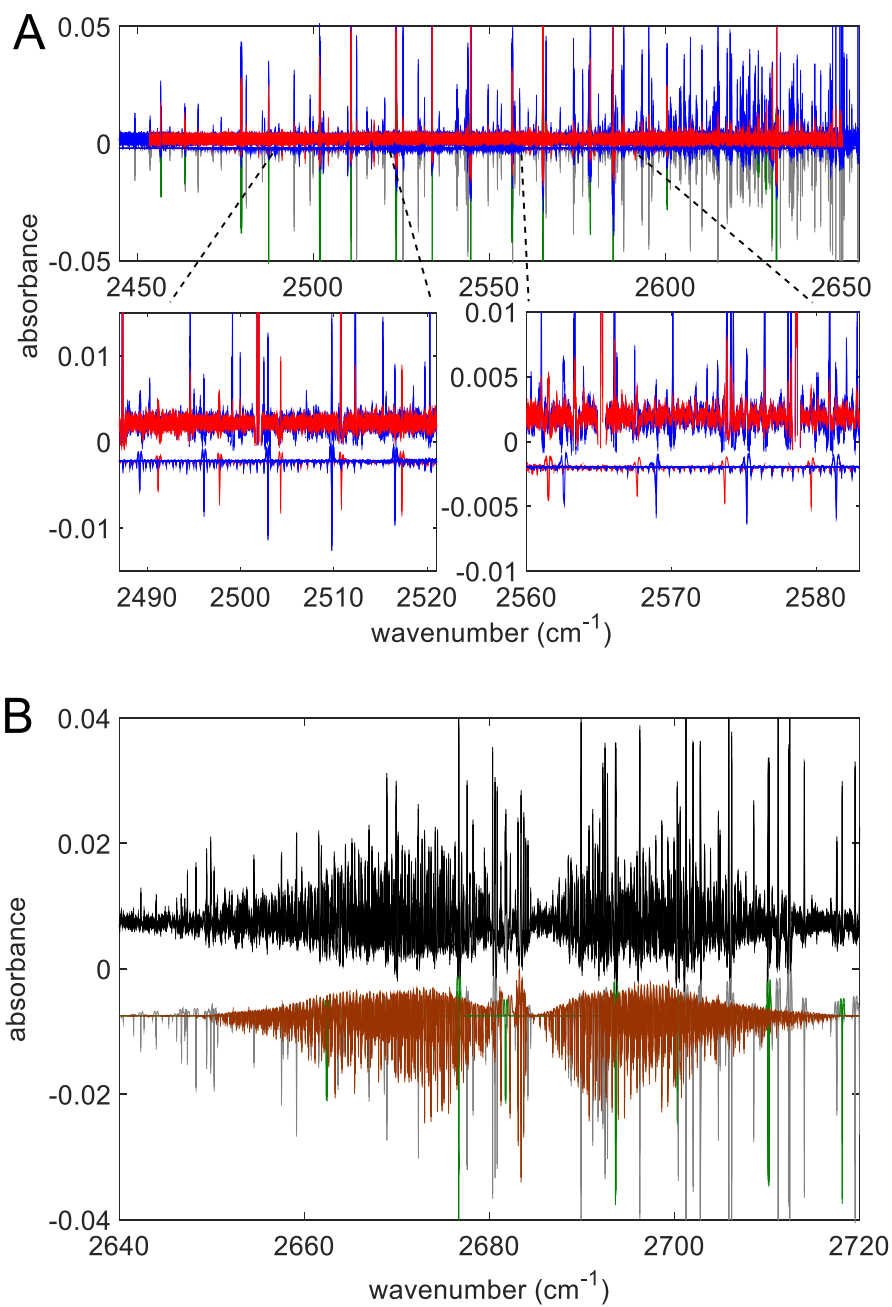

**Fig. 2**: (**A**) The full c*is*-DOCO spectrum (top). The data and simulation are given by positive and negative absorbance, respectively. The color codes are: *cis*-DO$^{12}$CO (blue), *cis*-DO$^{13}$CO (red), OD (green), $D_2O$ (gray). For clarity, only the simulated *cis*-DO$^{12}$CO and *cis*-DO$^{13}$CO are shown for comparison with experimental results in the zoomed-in bottom panels where the strongest Q branches are located. (**B**) The *trans*-DOCO spectrum (black). Simulated *trans*-DOCO (brown), OD (green), and $D_2O$ (gray) are shown as inverted absorbance. The detection spectral regions are OD (2400 to 2850 cm$^{-1}$), *cis*-DOCO (2450 to 2650 cm$^{-1}$) (Panel A), *trans*-DOCO (2640 to 2720 cm$^{-1}$) (Panel B), and $CO_2$ (2380 to 2445 cm$^{-1}$). Data in both panels were obtained by integrating from $t = 0$ to 200 μs at 100 μs camera integration time.

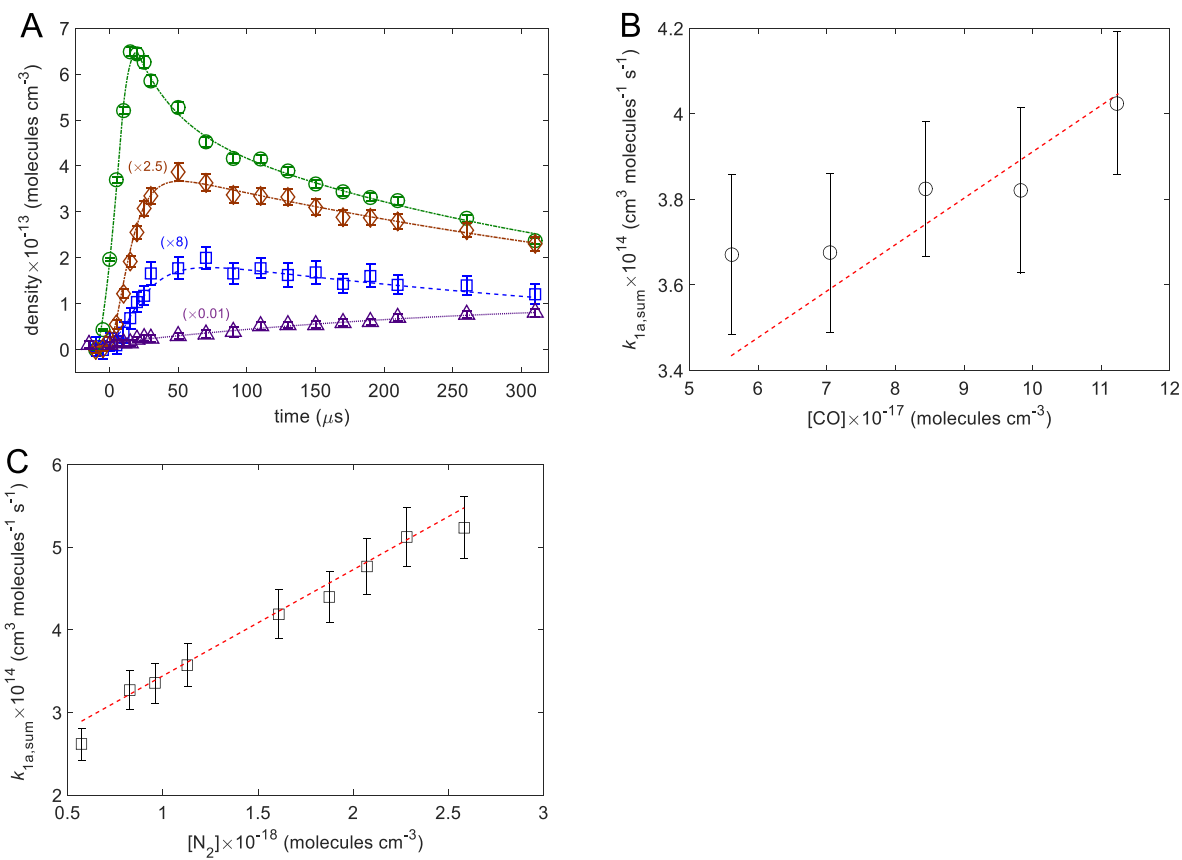

**Fig. 3**: (**A**) The time dependence of reactant OD (green circles), intermediates *cis*-DO$^{12}$CO (blue squares) and *trans*-DOCO (brown diamond), and product $CO_2$ (purple triangle) obtained from fitting full experimental spectra for a sequence of time windows after the initiation of the reaction. Quantities in parenthesis are multiplicative factors used to scale the traces to fit on a single plot. The camera integration time is 20 µs for OD, *cis*- and *trans*-DOCO, and 50 µs for $CO_2$. Each time point corresponds to ≈ 400 averaged spectra. The error bars are from uncertainties in the fit and the measured densities of the gases. The dashed and dotted lines are fit Eq. S4 for OD, Eq. 6 for *trans* and *cis*-DOCO, and Eq. S8 for $CO_2$ using values measured from this work. (**B**) $k_{1a,sum}$ is plotted as a function of CO while [$N_2$] is held constant at 8.9×10$^{17}$ molecules cm$^{-3}$. (**C**) $k_{1a,sum}$ is plotted as a function of $N_2$ while [CO] is held constant at 5.6×10$^{17}$ molecules cm$^{-3}$. In both plots, [$D_2$] and [$O_3$] are fixed at 1.4×10$^{17}$ and 2×10$^{15}$ molecules cm$^{-3}$, respectively. The red dashed lines show the fits obtained from a multidimensional linear regression procedure described in the main text. Data in panels B and C were obtained at 50 µs camera integration time. The error bars in panels B and C represent uncertainties from fits to Eqs. 8 to 9 and the measured densities of the gases.

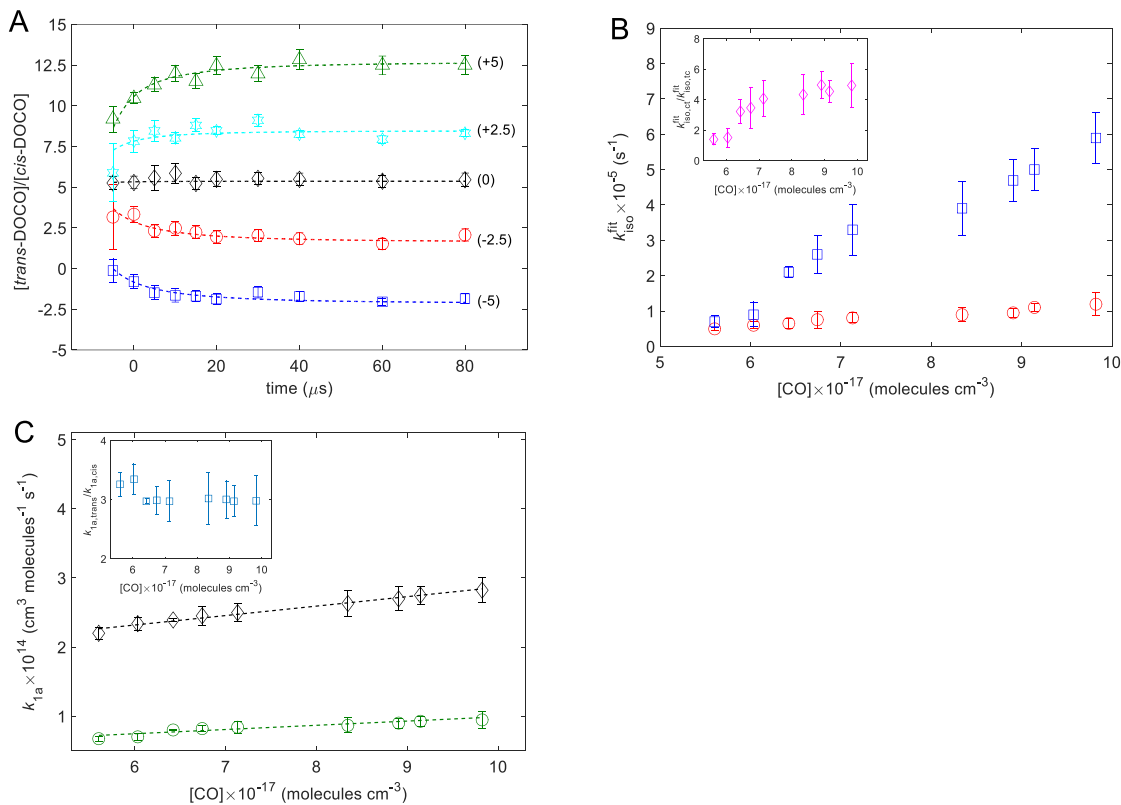

**Fig. 4:** (**A**) The ratio of [*trans*-DOCO] to [*cis*-DOCO] as a function of time for different CO concentrations. The CO concentrations are 5.6, 6.0, 6.4, 6.7, and 7.1×10$^{17}$ molecules cm$^3$ for blue squares, red circles, black diamonds, cyan hexagons, and green triangles, respectively. For these experiments, [N$_2$], [D$_2$] and [O$_3$] were fixed to 8.9×10$^{17}$, 1.4×10$^{17}$, and 2×10$^{15}$ molecules cm$^{-3}$, respectively. The arbitrary y-axis offsets of -5, -2.5, 0, +2.5, and +5, respectively, are added for ease of viewing. At high [CO] (> 6.4×10$^{17}$ molecules cm$^3$), all curves reach the same asymptotic value within experimental uncertainties. The dashed lines are fits of the rate equation model to the data. Data were obtained at 20 μs camera integration time. The error bars represent uncertainties from the spectral fit and measured densities of gases. (**B**) The fitted isomerization rate coefficients for *cis*→*trans* ($k_{iso,ct}^{fit}$, blue squares) and *trans*→*cis* ($k_{iso,tc}^{fit}$, red circles). The ratio of $k_{iso,ct}^{fit}$ / $k_{iso,tc}^{fit}$ as a function of [CO] is given in the inset. (**C**) The fitted effective bimolecular formation rate coefficients ($k_{1a}$) for *trans*-DOCO ($k_{1a,trans}$, black diamond) and *cis*-DOCO ($k_{1a,cis}$, green squares). The ratio of the formation rate coefficients $k_{1a,trans}$ / $k_{1a,cis}$ as a function of [CO] is given in the inset. The error bars from panels B and C represent uncertainties from the individual parameter variance and covariance between $k_{1a,sum}$ and $k_{iso}$.

Supplementary Material for

**Direct measurements of DOCO isomers in the kinetics of OD+CO**

**Authors:** T.Q. Bui[1]*, B.J. Bjork[1†], P.B. Changala[1], T.L. Nguyen[2], J.F. Stanton[2], M. Okumura[3], J. Ye[1]*

**Affiliations:**

[1]JILA, National Institute of Standards and Technology and University of Colorado, Department of Physics, University of Colorado, Boulder, CO 80309, USA

[2] Department of Chemistry, University of Florida, Gainesville, Florida 32611, USA

[3] Arthur Amos Noyes Laboratory of Chemical Physics, California Institute of Technology, Pasadena, California, 91125, USA

[†]Present address: Honeywell International, <u>303 Technology Ct., Broomfield, Colorado, 80021</u>, USA

*Correspondence to:  thbu8553@jila.colorado.edu, ye@jila.colorado.edu

**CONTENTS:**

Supplementary Materials

**Supplementary Materials**

§1. *cis-DOCO isotope shift and spectral parameters*

The experimental *cis*-DO$^{12}$CO and *cis*-DO$^{13}$CO spectra are presented in Fig. 2A. Compiled in Table S1 are the observed transitions and their approximate transition frequencies in wavenumber (cm$^{-1}$). Due to Doppler (~300 MHz) and instrumental (~900 MHz) broadening, we are not able to resolve the individual rotational features of each Q-branch.

**Table S1:** Experimental spectral features for *cis*-DO$^{12}$CO and *cis*-DO$^{13}$CO and their ro-vibrational assignments. The quantity in parenthesis provides an estimate of the uncertainty in the line position (in cm$^{-1}$) due to the FWHM linewidth of the unresolved Q-branch feature.

| $K_a'$ | $K_a''$ | $^{12}$C Position (cm$^{-1}$) | $^{13}$C Position (cm$^{-1}$) |
|---|---|---|---|
| 7 | 8 | 2489.220(0.147)* | 2491.106(0.147)* |
| 6 | 7 | 2496.040(0.08) | 2497.718(0.08) |
| 5 | 6 | 2502.931(0.05) | 2504.295(0.05) |
| 4 | 5 | 2509.764(0.05) | -- |
| 3 | 4 | 2516.524(0.05) | 2517.282(0.05) |
| 2 | 3 | -- | -- |
| 1 | 2 | -- | -- |
| 2 | 1 | 2548.102(N/A)* | 2547.572(N/A)* |
| 3 | 2 | -- | -- |
| 4 | 3 | 2562.550(0.13) | 2561.517(0.13) |
| 5 | 4 | 2568.954(0.11) | 2567.638(0.11) |
| 6 | 5 | 2575.191(0.10) | 2573.687(0.10) |
| 7 | 6 | 2581.414(0.10) | 2579.658(0.10) |
| 8 | 7 | -- | -- |

*not fitted
--transitions are too blended and/or SNR<2:1

By fitting the observed transitions to a Watson A-reduced asymmetric top Hamiltonian using PGOPHER, we obtain the vibrational band origin and A rotational constant (Table S2).

**Table S2:** Summary of spectral parameters for *cis*-DO$^{12}$CO and *cis*-DO$^{13}$CO. The average fit error is 0.014 cm$^{-1}$.

| | $^{12}$C(v=0) | $^{12}$C(v=1) | $^{13}$C(v=0) | $^{13}$C(v=1) |
|---|---|---|---|---|
| Origin (cm$^{-1}$) | 0 | 2539.909(3) | 0 | 2539.725(4) |
| A (MHz) | 110105.52[a] | 109313(4) | 106124(5) | 105423(5) |
| B (MHz) | 11423.441[a] | 11422.882[b] | 11420.075[c] | 11419.559[d] |
| C (MHz) | 10331.423[a] | 10324.228[b] | 10291.999[c] | 10284.951[d] |

[a]McCarthy *et al.* 2016
[b]$^{12}$C(v=0) value + VPT2 vibrational shifts
[c]$^{12}$C(v=0) value + VPT2 isotopic shift
[d]$^{12}$C(v=0) value + VPT2 isotopic shift & vibrational shift

**Table S3:** Comparison of experimental and theoretical isotopic shifts for *cis*-DOCO.

| | Experiment | Theory |
|---|---|---|
| Vibrational Shift (MHz) | | |
| $^{12}$C | -793(4) | -877.296 |
| $^{13}$C | -701(7) | -797.649 |
| Isotopic Shift (MHz) | | |
| v=0 | -3982(5) | -3976.926 |
| v=1 | -3890(6) | -3897.279 |
| isotopic shift in band origin (cm$^{-1}$) | -0.184 | -0.207[e] |
| | | -0.16[f] |

[e]VPT2
[f]variational calculation

§2 $k_{1a,sum}$ *fitting*

At early times ($t < 200$ μs), the *cis*- and *trans-DOCO* time dependence are described by the first-order differential equations,

$$\frac{d[cis]}{dt} = k_{1a,cis}\big[CO\big]\big[OD\big]_t - (k_{loss,cis}[X] + k_{iso,ct})[cis]_t + k_{iso,tc}[trans]_t$$
$$\frac{d[trans]}{dt} = k_{1a,trans}\big[CO\big]\big[OD\big]_t - (k_{loss,trans}[X] + k_{iso,tc})[trans]_t + k_{iso,ct}[cis]_t \tag{S1}$$

where $k_{iso,ct}$ and $k_{iso,tc}$ are the *cis→trans* and *trans→cis* isomerization rate coefficients, respectively. The subscript $t$ denotes time dependence. $[OD]_t$, $[cis]_t$, and $[trans]_t$ are the time-dependent concentrations of OD, *cis*-DOCO, and *trans*-DOCO, respectively, in the ground vibrational state. $k_{loss,cis}$ and $k_{loss,trans}$ are bimolecular loss rate coefficients for *cis*- and *trans*-DOCO, respectively, with one or more species X. By making the approximation that $k_{loss,cis} = k_{loss,trans} = k_{loss,sum}$, the rate equation for the sum of DOCO isomers is given by

$$\frac{d[DOCO_{sum}]}{dt} = k_{1a,sum}\big[CO\big]\big[OD\big]_t - k_{loss,sum}\big[DOCO_{sum}\big]_t\big[X\big]. \tag{S2}$$

Here, $[DOCO_{sum}] \equiv [trans]+[cis]$ and $k_{1a,sum} \equiv k_{1a,cis}+k_{1a,trans}$. The isomerization terms from Eq. S1 cancel out when only the total concentration of DOCO is considered. The Laplace transform of Eq. S2 gives the solution in Eq. S3, which is the convolution of DOCO loss term with $[OD]_t$.

$$[DOCO_{sum}]_t = k_{1a,sum}[CO]\int_0^t e^{-(k_{loss,sum}[X])(t-u)}[OD](u)du . \tag{S3}$$

Here, $u$ is a dummy variable. To fit the data, we use the empirical functional form of $[OD]_t$ comprised of sum of exponential rise and fall components.

$$[OD]_t = a_1 e^{-b_1 t} + a_2 e^{-b_2 t} - (a_1 + a_2)e^{-b_3 t} , \tag{S4}$$

Here, $b_1$ and $b_2$ are bi-exponential decay terms while $b_3$ is a rise term. This procedure is equivalent to a spline interpolation in which $a_1$, $a_2$, $b_1$, $b_2$, and $b_3$ are fitted independently. Using Eq. S4, Eq. S3 is given by

$$[DOCO_{sum}]_t = k_{1a,sum}[CO]\left( a_1 \frac{e^{-b_1 t} - e^{-r_{loss,sum} t}}{b_1 - r_{loss,sum}} + a_2 \frac{e^{-b_2 t} - e^{-r_{loss,sum} t}}{b_2 - r_{loss,sum}} - (a_1 + a_2)\frac{e^{-b_3 t} - e^{-r_{loss,sum} t}}{b_3 - r_{loss,sum}} \right). \tag{S5}$$

We fit Eqs. S4 and S5 to the measured $[OD]_t$ and $[DOCO]_t$ to obtain the $DOCO_{sum}$ formation and loss rate coefficients, $k_{1a,sum}$ and $n_{loss,sum}$, respectively. From our fitted values of the bimolecular rate coefficients $k_{1a,sum}$, we determined the termolecular rate coefficients from a multidimensional linear regression to the expression

$$k_{1a,sum} = k_{1a,sum}^{(CO)}[CO] + k_{1a,sum}^{(N2)}[N_2] + k_{1a,sum}^{(D2)}[D_2] + k_{1a,sum}^{(0)}. \tag{S6}$$

The rate equation for the OD time dependence is:

$$\frac{d[OD]}{dt} = k_{D2}[D_2]\left[O(^1D)\right]_t + k_{O3}[O_3][D]_t - (k_{1a,cis} + k_{1a,trans} + k_{1b})[CO][OD]_t. \tag{S7}$$

Here, the $k_{D2}$ ($1.2\times10^{-10}$ cm$^3$ molecules$^{-1}$ s$^{-1}$) and $k_{O3}$ ($2.9\times10^{-11}$ cm$^3$ molecules$^{-1}$ s$^{-1}$) bimolecular rate coefficients to form OD are assumed to be equivalent to their hydrogen analogues obtained from the JPL Chemical Kinetics and Photochemical Data evaluation.

The rate equation for the $CO_2$ time dependence is

$$\frac{d[CO_2]}{dt} = k_{1b}[CO][OD]_t - k_{pump}[CO_2]_t. \tag{S8}$$

The only loss channel for $CO_2$ is the gas pump out time ($k_{pump}$) of the cavity, which is approximately 20 ms.

*OD+CO Chemical Mechanism:*

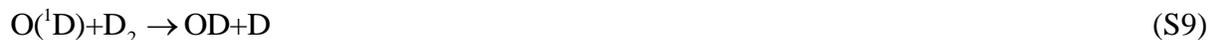
$$O(^1D)+D_2 \rightarrow OD+D \tag{S9}$$

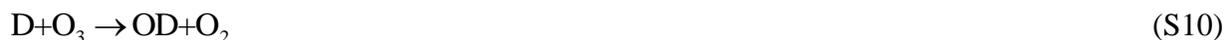
$$D+O_3 \rightarrow OD+O_2 \tag{S10}$$

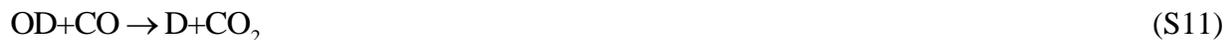
$$OD+CO \rightarrow D+CO_2 \tag{S11}$$

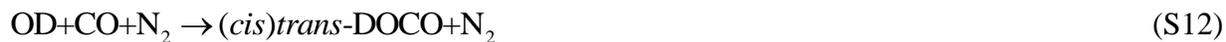
$$OD+CO+N_2 \rightarrow (cis)trans\text{-}DOCO+N_2 \tag{S12}$$

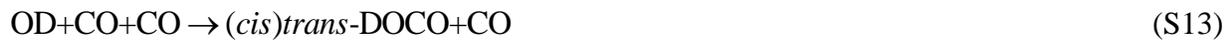
$$OD+CO+CO \rightarrow (cis)trans\text{-}DOCO+CO \tag{S13}$$

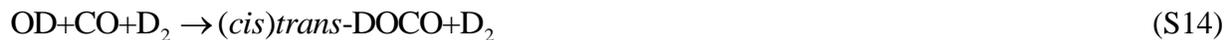
$$OD+CO+D_2 \rightarrow (cis)trans\text{-}DOCO+D_2 \tag{S14}$$

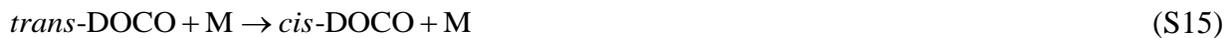
$$trans\text{-}DOCO + M \rightarrow cis\text{-}DOCO + M \tag{S15}$$

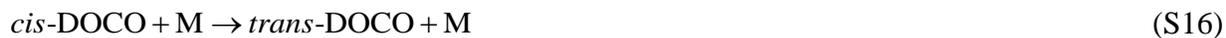
$$cis\text{-}DOCO + M \rightarrow trans\text{-}DOCO + M \tag{S16}$$

(*cis*)*trans*-DOCO + X → products                                                    (S17)

Notes:

1) OD are formed in steady-state, especially at longer times. Since we can directly observe OD, we do not need to account for its loss/removal in this scheme.

2) We measure the kinetics of DOCO formation at early times, to minimize biases introduced by secondary reactions, especially radical-radical reactions involving DOCO.

*Effect of O₃:* In order to determine the DOCO+O₃ reaction rate, we perform a global fit of $k_{1a,sum}$ and $r_{loss,sum}$ across all of the CO, N₂, D₂, and O₃ data sets. From this procedure, we obtain an average O₃ loss rate coefficient $k_{loss,sum} = (2.5\pm0.6)\times10^{-11}$ cm³ molecules⁻¹ s⁻¹ (Fig. S1). The total loss rate, $r_{loss,sum}$, was fixed to this value for the final determination of $k_{1a,sum}$. Losses may be isomer specific but are not resolvable with measurement of the total loss. In addition, thermalized *cis*-DOCO loss to D+CO₂ (via tunneling) is expected to be much slower than reaction with O₃.

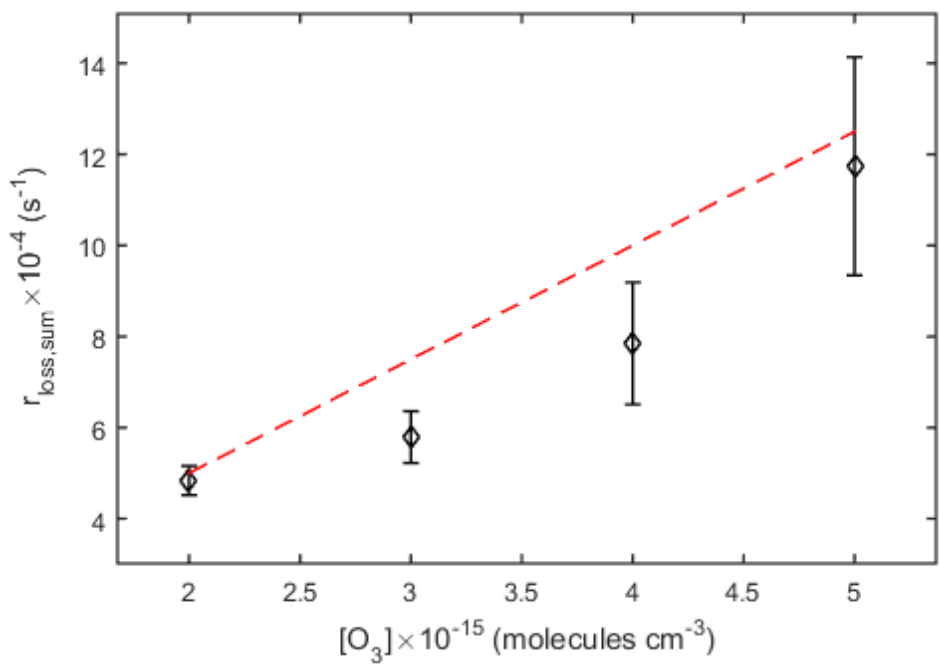

**Fig. S1.** Variation of $r_{loss,sum}$ with O₃ concentration.

*Effect of $D_2$:* To determine the impact of $D_2$ gas on $k_{1a,sum}$, we vary $D_2$ under constant $[N_2]=9.8\times10^{17}$, $[CO]=5.6\times10^{17}$, and $[O_3]=2\times10^{15}$ molecules cm$^{-3}$. The results are shown in Fig. S2. No statistically significant variation with $D_2$ is observed. Nonetheless, the variation of $k_{1a,sum}$ with $D_2$ was included in the multidimensional linear regression, with a fitted rate coefficient of $k_{1a,sum}^{(D2)} = 9.6_{-24}^{23}\times10^{-33}$ (cm$^6$ molecules$^{-2}$ s$^{-1}$).

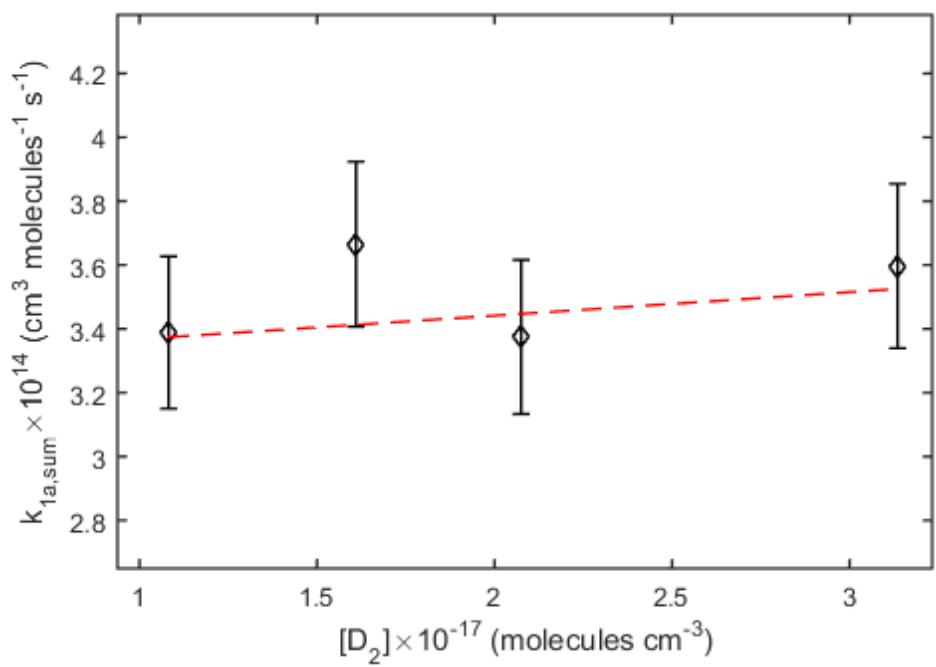

**Fig. S2.** Variation of $k_{1a,sum}$ with $D_2$ concentration.

**Table S4:** Summary of statistical (stat) and systematic (sys) errors for $k_{1a,sum}$.

| | | Error Source | $k_{1a,sum}^{(N2)}$ | $k_{1a,sum}^{(CO)}$ |
|---|---|---|---|---|
| **$k_{1a,sum}$ ($cm^6$ molecules$^{-2}$ s$^{-1}$)** | | | $1.3\times10^{-32}$ | $8.5\times10^{-33}$ |
| **Statistical Errors** | | (statistical, fit residual) | 15% | 47% |
| **Experimental Control** | §1 | Flow & Pressure | 7% (stat) | |
| **Molecular Parameters** | §2 | OD Cross Section | 10% (stat) | |
| | §2 | *trans*-DOCO Cross Section | 10% (stat) | |
| | §2 | *cis*-DOCO Cross Section | 20% (stat) | |
| **Data Analysis** | | Cross-contamination of OD and $D_2O$ | -1% (sys) | |
| | | Total Systematic Error | (-11%,+0%) | |
| | | Total Statistical Error | 30%, 53% | |
| | | Total Error Budget | (-41%,+30%) | (-64%,+53%) |

**Table S5**: Summary of fitted rate coefficients

| | |
|---|---|
| **$k_{1a,sum}^{(CO)}$ ($cm^6$ molecules$^{-2}$ s$^{-1}$)** | $8.5_{-5.4}^{4.5}\times10^{-33}$ |
| **$k_{1a,sum}^{(N2)}$ ($cm^6$ molecules$^{-2}$ s$^{-1}$)** | $1.3_{-0.5}^{0.4}\times10^{-32}$ |
| **$k_{1a,sum}^{(D2)}$ ($cm^6$ molecules$^{-2}$ s$^{-1}$)** | $9.6_{-24}^{23}\times10^{-33}$ |
| **$k_{1b}^{g}$ ($cm^3$ molecules$^{-1}$ s$^{-1}$)** | $5.6(7)\times10^{-14}$ |
| **$k_{loss,sum}$ ($cm^3$ molecules$^{-1}$ s$^{-1}$)** | $2.5(6)\times10^{-11}$ |
| **$k_{1a,trans}^{(CO)}$ ($cm^6$ molecules$^{-2}$ s$^{-1}$)** | $1.4(4)\times10^{-32}$ |
| **$k_{1a,cis}^{(CO)}$ ($cm^6$ molecules$^{-2}$ s$^{-1}$)** | $6(2)\times10^{-33}$ |
| **$k_{iso,tc}^{(CO)}$ ($cm^3$ molecules$^{-1}$ s$^{-1}$)** | $1.3(2)\times10^{-13}$ |
| **$k_{iso,ct}^{(CO)}$ ($cm^3$ molecules$^{-1}$ s$^{-1}$)** | $7.5(2)\times10^{-13}$ |

[g]Bui *et al.* 2017

## §3. Theoretical calculations for OD+CO

*Theoretical rate coefficients for DOCO isomerization:* Using a theoretical technique that combines semi-classical transition state theory (SCTST) with the 2-dimensional master-equations (2DME), we calculate the pressure-dependent *cis→trans* and *trans→cis* isomerization rate coefficients, denoted as $k_{iso,ct}$ and $k_{iso,tc}$, respectively. This method has previously provided accurate predictions of the pressure-dependent $k_1(T,P)$ for the OH(OD)+CO reaction. The chemical kinetics analysis is performed using the potential energy surface (PES, Fig. 1) calculated using the HEAT protocol. The third-body collisional partner is air ($N_2$ = 80%, $O_2$=20%). The results are shown in Fig. S3. The theoretical equilibrium constant for DOCO isomerization is defined as $K_{iso,th} = k_{iso,ct}/k_{iso,tc} \approx 14:1$.

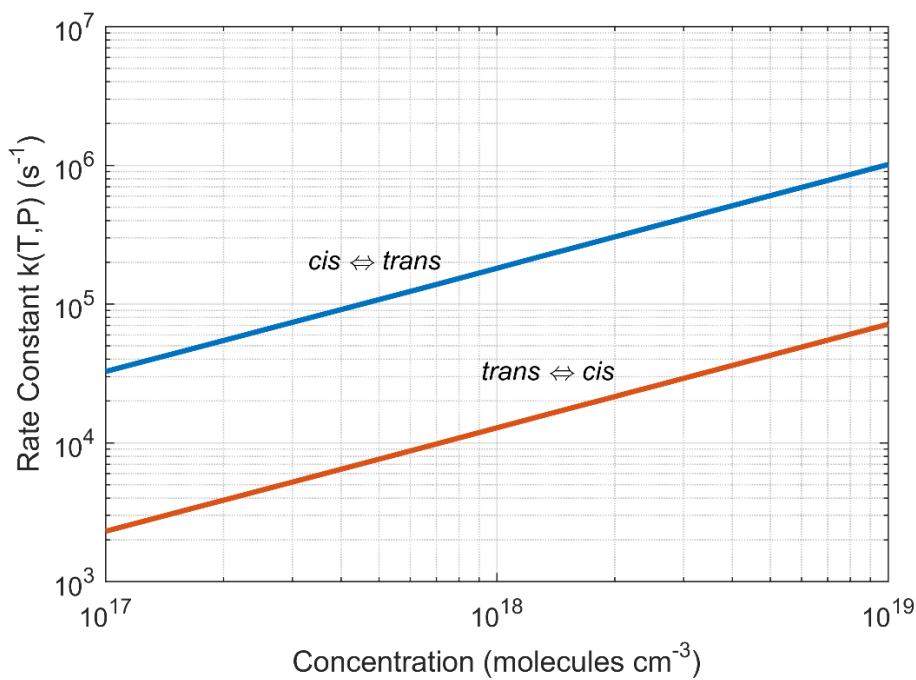

**Figure S3:** Calculated *cis/trans* isomerization rates for DOCO isomers using semi-classical transition state theory (SCTST) and 2D master-equation. The calculated k(*T, P*) for isomerization (forward and backward) have slopes of less than unity and thus exhibit characteristic "falloff" behavior.

In the low-pressure regime, $k_1 = k_{1a,cis} + k_{1a,trans} + k_{1b}$, where $k_{1b}$ is the bimolecular rate coefficient for the $D+CO_2$ channel. Using $k_{1a,cis}$ and $k_{1a,trans}$ measured in this work, $k_1$ as a function of $N_2$ pressure is plotted in green (1σ uncertainty) in Fig. S4. The results are in good agreement with $k_1$ measurements for OD+CO from Golden *et al*. and Paraskevopoulos & Irwin.

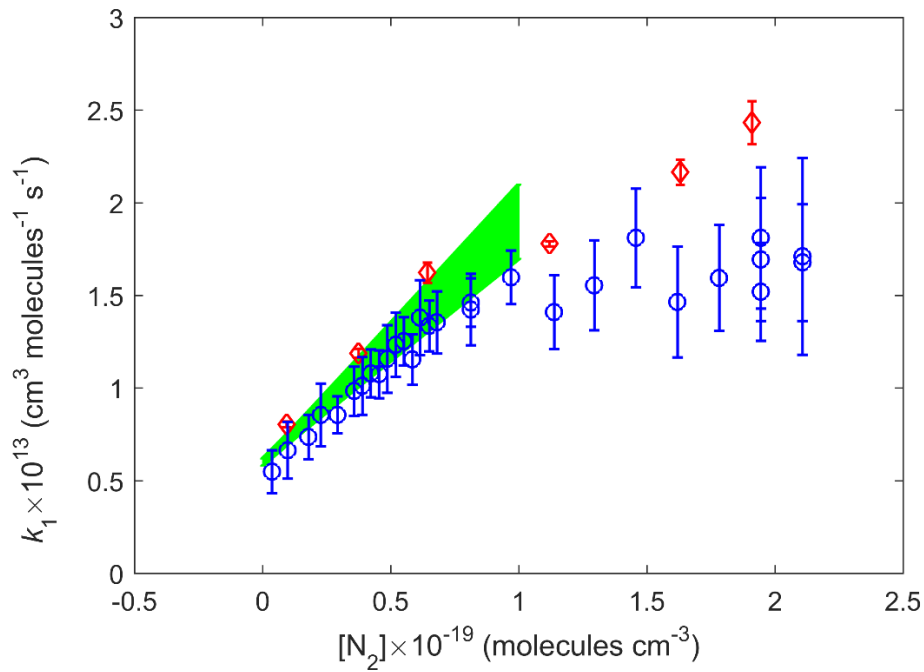

**Fig. S4**: Our measured $k_1$ comparison with previous works. The measured $k_1$ as a function of $N_2$ are given by the shaded green region (within 1σ uncertainties). Red diamond and blue circles are the measured $k_1$ values from Golden *et al*. and Paraskevopoulos & Irwin, respectively.